\documentclass{llncs}
\pdfoutput=1
\usepackage[english]{babel}
\usepackage{float}
\usepackage{graphicx,epstopdf}
\usepackage{amsmath,amssymb,amsfonts,subfigure}
\usepackage{comment}
\usepackage{cite}
\usepackage{adjustbox}

\title{Estimating the Degree Centrality Ranking of a Node}
\author{Akrati Saxena, Vaibhav Malik, S. R. S. Iyengar\\
\normalsize{\{akrati.saxena, vaibhav.malik, sudarshan\}@iitrpr.ac.in}\\
}
\institute{Department of Computer Science,\\
\normalsize{Indian Institute of Technology Ropar, India}}

\begin{document}
\maketitle

\begin{abstract}
Complex networks have gained more attention from the last few years. The size of real-world complex networks, such as online social networks, WWW network, collaboration networks, is increasing exponentially with time. It is not feasible to collect the complete data and store and process it. In the present work, we propose a method to estimate the degree centrality rank of a node without having the complete structure of the graph. The proposed algorithm uses the degree of a node and power-law exponent of the degree distribution to calculate the ranking. Simulation results on the Barabasi-Albert networks show that the average error in the estimated ranking is approximately $5\%$ of the total number of nodes.
\end{abstract}
\section{Introduction}


Properties of a complex network can be categorized at the network level called macro-level properties as well as at the node level called micro-level properties. Examples of the major macro-level properties are small-world phenomenon (six degree separation) \cite{milgram1967small, travers1969experimental, newman2000models}, scale-free degree distribution \cite{barabasi1999emergence, caldarelli2007scale}, preferential attachment \cite{newman2001clustering}, global clustering coefficient \cite{newman2003structure, klemm2002highly} etc. Except for these properties, each node also has some unique characteristics that distinguish it from other nodes, called micro-level properties. An example of such properties is centrality measures. Centrality measures are used to quantify the importance of nodes in the network under different contexts. Different centrality measures have been coined to suit different applications. There are some centrality measures that can be calculated using local information of the node, like degree centrality \cite{shaw1954some}, and semi-local centrality measure \cite{chen2012identifying}. Others use global information of the network like closeness centrality \cite{sabidussi1966centrality}, betweenness centrality \cite{freeman1977set}, eigenvector centrality \cite{stephenson1989rethinking}, Katz centrality \cite{katz1953new} and so on. 

Real-world complex networks are large-scale and dynamic in nature. So, the calculation of different centrality measures based on the global structure of the network is very costly. Fast algorithms to update the values of different centrality measures for dynamic graphs have been proposed, such as for closeness centrality \cite{sariyuce2013incremental, yen2013efficient, kas2013incremental}, betweenness centrality \cite{green2012fast, lee2012qube, goel2014faster, nasre2014betweenness}, PageRank \cite{desikan2005incremental, bahmani2012pagerank, berberich2007comparing}. Another important point to consider is that the network size also increases exponentially with time. It is hard to collect and store the complete data on a single system and process it. This motivated researchers to propose complex network algorithms based on the local information. Many approximation and heuristic algorithms have been proposed to get good approximated values for different centrality measures \cite{bader2007approximating, eppstein2004fast, geisberger2008better}. These algorithms use sampling techniques, such as uniform sampling, random walk, snowball sampling, weighted sampling. Lee et al. studied different properties of the network, such as power-law exponent of the degree distribution, clustering coefficient, assortativity, average path length, and betweenness centrality using three different sampling techniques\cite{lee2006statistical}. Goldreich et al. presented an estimator to calculate the average degree using uniform sampling \cite{goldreich2008approximating}. In real-world networks, uniform sampling is not always feasible. This leads to a motivated study towards other sampling methods using random walks \cite{lovasz1993random, cooper2009random}. Kurant et al. proposed the Induced Edges (IE) technique based on the random walk to estimate the total number of nodes \cite{kurant2012graph}. Hardiman et al. estimated the clustering coefficient and the total number of nodes using a random walk, that is implemented using public interface call \cite{hardiman2013estimating}. Similar techniques to calculate the total number of nodes in exponentially growing networks are studied by \cite{katzir2011estimating, cooper2014estimating}. Dasgupta et al. calculate the average degree of the network by taking $O(log k_{max} \cdot log log k_{max})$ samples, where $k_{max}$ is an upper bound on the maximum degree of the nodes\cite{dasgupta2014estimating}. All these techniques are storage efficient, as we only collect some data samples and store them. 

As we discussed, current literature focuses mainly on approximating values of different centrality measures or network parameters. However, in real-life applications, most of the time, the actual value is not important, \textit{what's important is where you stand}, not with respect to the mean but with respect to others. For example, in entrance examination systems, importance is given to the percentile of a person, not the percentage. Percentile is an estimate of the ranking of a candidate. Similarly, in social networks, each person would like to estimate her ranking to know how strong she is in the network. This ranking can be based on any centrality measure or network parameter depending on the context. Fortunato et al. proposed an algorithm to approximate PageRank value using in-degree of the node \cite{fortunato2005make}. Calculating global parameters using local information of the node is the future of the approximation algorithms in complex networks \cite{saxena2017global}. In this paper, we focus on estimating the degree centrality ranking of a node using local information. A node having more number of neighbors has a higher ranking.  

A straightforward way to calculate the degree centrality rank of a node, is to get the degree centrality value of all the nodes, order them, and get the rank of the desired node. The complexity of this process will be $O(n log n)$, given that we have the entire network in hand. Instead of this, we propose a method to estimate the degree ranking of a node using network parameters without having the complete structure of the graph. In real-world networks, degree centrality follows a power-law distribution. Barabasi and Albert observed this pattern and proposed an evolving model (BA networks) to generate synthetic networks similar to real-world networks \cite{barabasi1999emergence}. This model is based on preferential attachment where probability of a node getting a new connection is directly proportional to its degree.

The present work estimates degree centrality ranking of a node in BA networks. We use sampling-based techniques to calculate different network parameters, such as the total number of nodes, average, and minimum degree in the network. These parameters and power-law exponent of the degree distribution is used to estimate the ranking of a node. We also study variance for the proposed ranking method. Simulation results are shown on the BA networks and real-world networks. It is shown that there is a small error in the estimated ranking. Error is calculated, as the modular difference of the actual and estimated ranking. We have also proposed fast rank estimation methods for other centrality measures, such as sampling-based methods for degree ranking \cite{saxena2017observe, saxena2018estimating}, heuristic methods for closeness ranking \cite{saxena2017fast, saxena2019heuristic}, and k-shell index \cite{saxena2018k}.

The rest of this paper is organized as follows. Section 2 describes the properties of the BA model followed by the mathematical analysis of the degree centrality ranking. Section 3 contains experimental results, and section 4 concludes the paper along with future directions.

\section{Model and Background}

In this section, we introduce the method based on the BA model to estimate the degree centrality ranking of a node. A graph is represented as $G(V,E)$, where $V$ is a set of nodes, and $E$ is a set of edges. $n$ is the total number of nodes, and $m$ is the total number of edges in the network. Minimum, maximum, and average degree of the nodes is represented by $k_{min}$, $k_{max}$, and $d_{avg}$ respectively. All these parameters and the total number of nodes are calculated using sampling-based methods \cite{dasgupta2014estimating}. 
Degree centrality of a node $u$ is denoted by $deg(u)$, that represents total number of neighbors of the node. Degree centrality ranking of a node $u$ is defined as, $r(u) = \sum_{\forall v \neq u} X_{vu}+1$ , where $X_{vu}$ is indicator random variable whose value is 1 when $deg(v) > deg(u)$ and 0 otherwise.
Next, we discuss the preferential attachment and power-law degree distribution proposed by the BA model.

\subsection{BA Model}
Barabasi and Albert proposed an evolutionary preferential attachment model for the formation of real-world complex networks \cite{barabasi1999emergence}. This model starts with a seed graph that contains $n_0$ disconnected nodes. At each timestamp, a new node is added, and it is connected with $m$ already existing nodes. The probability $\prod (deg(x))$ of an existing node $x$ to get new connection depends on the degree $k_x$ of node $x$. It is defined as,

\begin{center}
$\prod (deg(x)) = \frac{deg(x)}{\sum_{y}deg(y)}$
\end{center}

So, the nodes having higher degrees acquire more links over time, thereby skewing the distribution towards lower degree. Preferential attachment model gives rise to power law degree distributions, where probability of a node having degree $k$ is $f(k)$, that is defined as, 

\begin{center}\begin{equation} \label{eq:fk}
f(k) = ck^{-\gamma}
\end{equation}\end{center}

where, $\gamma$ is the power exponent and for real world networks its range is $2 \leq \gamma \leq 3$. As the network grows, only a few nodes, called hubs, manage to get a large number of links. 

\subsection{Degree Ranking}
In this section, we propose the mathematical method to calculate the degree centrality ranking in complex networks.

Consider graph $G(V,E)$, where $|V|$  is $n$, in which each node $u \in V$ picks its degree $deg(u)$ from the power law distribution given in equation \eqref{eq:eq2}, independent of the degrees of the other nodes, i.e. probability that degree of node $u$ is $k$, is defined as,
\begin{center}\begin{equation} \label{eq:eq2}
P(deg(u) = k) = \int_{k - 1/2 }^{k + 1/2 }f(k) 
\end{equation}\end{center}
where,  $f(k)$ is power law function from \eqref{eq:fk}. Here $k$ varies from $k_{min}$ to $k_{max}$, where $k_{min}$ and $k_{max}$ are minimum and maximum degree of a node. Let us define $p$ as probability of a node having degree greater than k, i.e.
\begin{center}
$p = P(deg(u) > k) = \int_{k + 1/2 }^{k_{max} + 1/2 }f(k) dk$
\end{center}
\begin{center}
$p = P(deg(u) > k) = \int_{k + 1/2 }^{k_{max} + 1/2 } ck^{-\gamma} dk$
\end{center}
To calculate $p$, first we derive $c$ and $\gamma$. By axiom of probability, we know that integrating  $f(k)$ from $k_{min}$ to $k_{max}$ will be equal to 1, i.e.

\begin{center}
$\int_{k_{min} - 1/2 }^{k_{max} + 1/2 }f(k) = 1 $
\end{center}
After integrating,
\begin{center}

$c \frac{ (k_{min} - 1/2)^{1-\gamma} - (k_{max} + 1/2)^{1-\gamma} }{ \gamma - 1 } = 1$
\end{center}
\begin{center}
$c = \frac{ \gamma - 1 }{ (k_{min} - 1/2)^{1-\gamma} - (k_{max} + 1/2)^{1-\gamma} }$
\end{center}
To calculate $\gamma$, we will use average degree $d_{avg}$. Using the estimator proposed by A Dasgupta et al. \cite{dasgupta2014estimating}, we can estimate the average degree of a graph. By using average degree of the graph, we will calculate the power law exponent $\gamma$ of the degree distribution. By definition average degree degree of a graph can be written as,

\begin{center}
$d_{avg} = \int_{k_{min} - 1/2}^{k_{max} + 1/2} kf(k) $
\end{center}
\begin{center}
$d_{avg} = \int_{k_{min} }^{k_{max} } kck^{-\gamma} $
\end{center}
After Integrating,
\begin{center}
$d_{avg} = c \frac{(k_{min} - 1/2) ^{2-\gamma} - (k_{max} + 1/2)^{2-\gamma}}{\gamma-2} $
\end{center}
Using value of $c$ in the above equation,
\begin{center}
$d_{avg} = \frac{\gamma-1}{\gamma-2} \cdot \frac{ (k_{min}-1/2)^{2-\gamma} - (k_{max}+1/2)^{2-\gamma}}{ (k_{min}-1/2)^{1-\gamma} - (k_{max}+1/2)^{1-\gamma} } $
\end{center}
$\gamma > 2$ for most of the real world networks and BA model,
So as $n \rightarrow \infty$,
\begin{center}
$(k_{max}+1/2)^{2-\gamma} \rightarrow 0$
\end{center}
and,
\begin{center}
$(k_{max}+1/2)^{1-\gamma} \rightarrow 0$
\end{center}

Therefore the equation can be written as,
\begin{center}
$d_{avg} = \frac{\gamma-1}{\gamma-2} (k_{min}-1/2)$
\end{center}
\begin{center}
i.e. $ \gamma = 2 + \frac{k_{min}-1/2}{d_{avg}-k_{min}+1/2}$
\end{center}
Now,
\begin{center}
$p = P(deg(u) > k) = \int_{k + 1/2 }^{k_{max} + 1/2 } ck^{-\gamma} dk$
\end{center}
\begin{center}

$p = c \frac{ (k + 1/2)^{1-\gamma} - (k_{max} + 1/2)^{1-\gamma} }{ \gamma - 1 }$
\end{center}
using value of $c$, the equation will be,
\begin{center}
$p = \frac{ \gamma - 1 }{ (k_{min} - 1/2)^{1-\gamma} - (k_{max} + 1/2)^{1-\gamma} } \frac{ (k + 1/2)^{1-\gamma} - (k_{max} + 1/2)^{1-\gamma} }{ \gamma - 1 }$
\end{center}
\begin{center}

$p \approx \left (  \frac{k+1/2}{k_{min}-1/2}\right )^{1-\gamma}$

\end{center}
In the given network, each node will choose its degree according to the distribution as given by equation \eqref{eq:eq2}. Let say a node $v$ has a degree $k$. So, probability that the rank of this node is $\alpha$, given that its degree is $k$, will be,
As we know, each node chooses its degree from the distribution given in equation \eqref{eq:fk}.  
\begin{center}
$P(r(v)=\alpha \mid deg(v)=k) = \begin{pmatrix}n-1\\  \alpha-1\end{pmatrix} p^{\alpha-1} (1-p)^{n-\alpha}$
\end{center}

i.e. rank of node $v$ will be $\alpha$, when $\alpha - 1$ nodes have their degree to be greater than k. Rank of a node can vary from $1$ to $n$, so we calculate expected value of the ranking of a node with degree k. It can be defined as,
\begin{center}
$E[r(v) \mid deg(v)=k] = \sum_{\alpha = 1}^{n} \alpha P(r(v)=\alpha \mid d(v)=k)$
\end{center}
\begin{center}

$=\sum_{\alpha = 1}^{n} \alpha \begin{pmatrix}n-1\\  \alpha-1\end{pmatrix} p^{\alpha-1} (1-p)^{n-\alpha}$ 
\end{center}
Solution of the above equation will provide us expected ranking of the node,
\begin{center}
$E[r(v) \mid deg(v)=k] = (n-1)p + 1$
\end{center}
Now, we calculate the variance of the ranking of a degree $k$ node,
\begin{center}
$Var(r(v) \mid deg(v)=k) = E[r(v)^2 \mid deg(v)=k] - E[r(v) \mid deg(v)=k]^2$
\end{center}
\begin{center}
$E[r(v)^2 \mid deg(v)=k] = \sum_{\alpha = 1}^{n} \alpha^2 P(r(v)=\alpha \mid d(v)=k)$
\end {center}
On solving it,
\begin{center}
$E[r(v)^2 \mid deg(v)=k] = (n-1)p(1-p) + (n-1)^2p^2 + 2(n-1)p$
\end {center}
By putting the expected value, variance can be written as,
\begin{center}
$Var(r(v) \mid deg(v)=k) = (n-1)p - (n-1)p^2 -2$
\end{center}
Real world complex networks are known to have more nodes having smaller degree. We can see from the variance formula that the nodes having higher value of $p$ will have high variance in the degree rank estimation. Similarly, the nodes having higher degree have smaller probability value, so there will be less variance and the expected rank will be closer to the actual rank of the node.

\section{Simulation Results}

We create synthetic networks of different sizes using the BA model, where, each new coming node makes $m=10$ connections. To verify the proposed method of degree centrality ranking, we plot the graphs of actual and expected degree ranking for the generated networks. The actual ranking of a node is calculated by sorting all the nodes in the given graph as per their degree and then assign them ranking. The expected ranking is calculated using the proposed method. According to the model, the highest degree node has to rank $1$. 

In Fig. 1, actual, and expected rankings are plotted using red and blue color dots. Two thick lines drawn above and below the rankings are plotted to show the calculated variance for the proposed method. It can be observed in Fig. 1, that prediction for the higher degree nodes is more accurate than the lower degree nodes. Similarly, higher degree nodes have less variance than lower degree nodes. All the expected and actual rankings lie between the proposed range of variance. There is some difference between the actual and expected ranking. It occurs because we integrate to calculate the ranking by assuming that the nodes of all degrees (from $k_{min}$ to $k_{max}$) are present, but it does not happen in the complex networks. There can be missing a few degrees in the degree distribution; for example, there can not be any node of degree 31 even if there exist nodes of degree 30 and 32. 

We also study the average difference in the expected ranking and actual ranking of the nodes that is calculated by taking the modular difference of these two rankings. Table 1 shows the average error for different size networks. If we do not consider all nodes having the smallest few degrees (around the smallest 15 degrees) while calculating the variance, then it is very less. It shows that the proposed method has a very small error for the higher degree nodes, and their ranking can be predicted with high accuracy.

\begin{figure}
     \begin{center}
        \subfigure[]{%
            \label{fig:first}
            \includegraphics[width=0.5\textwidth]{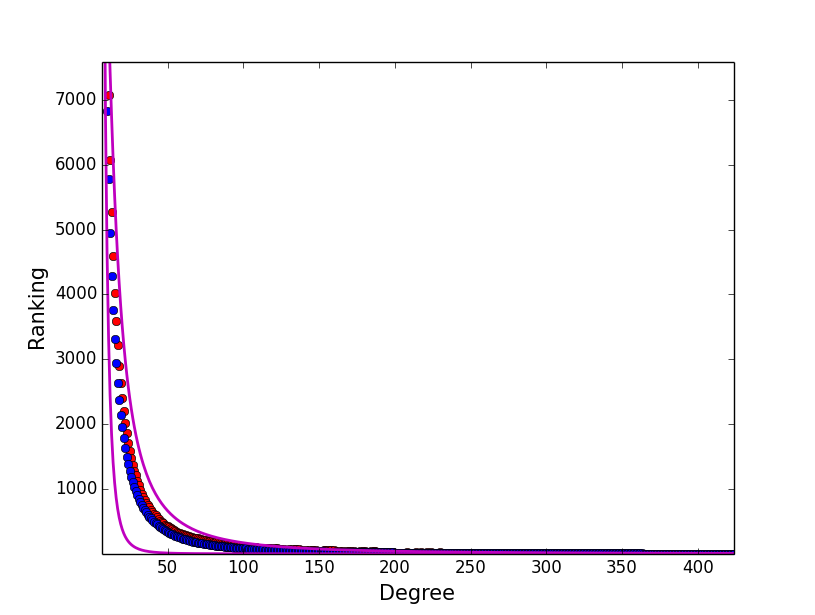}
        }%
        \subfigure[]{%
           \label{fig:second}
           \includegraphics[width=0.5\textwidth]{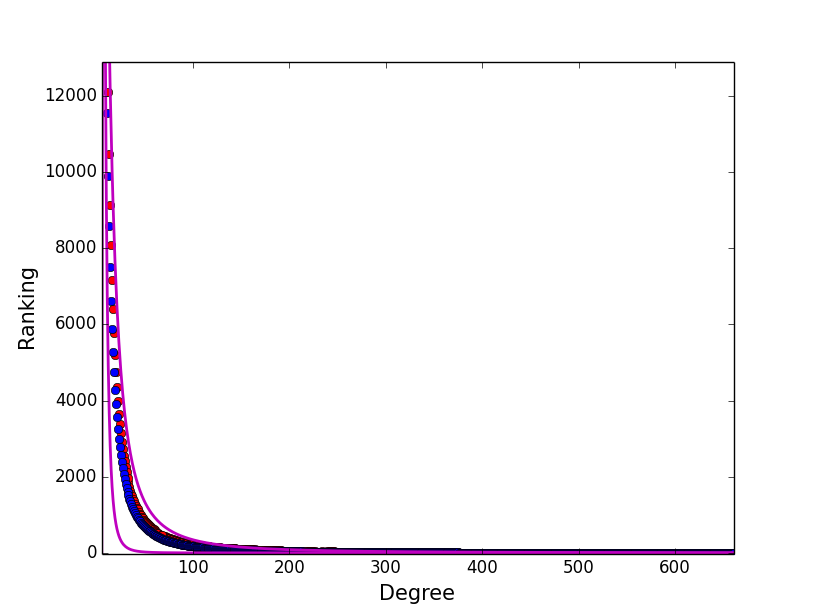}
        }\\ 
        \subfigure[]{%
            \label{fig:third}
            \includegraphics[width=0.5\textwidth]{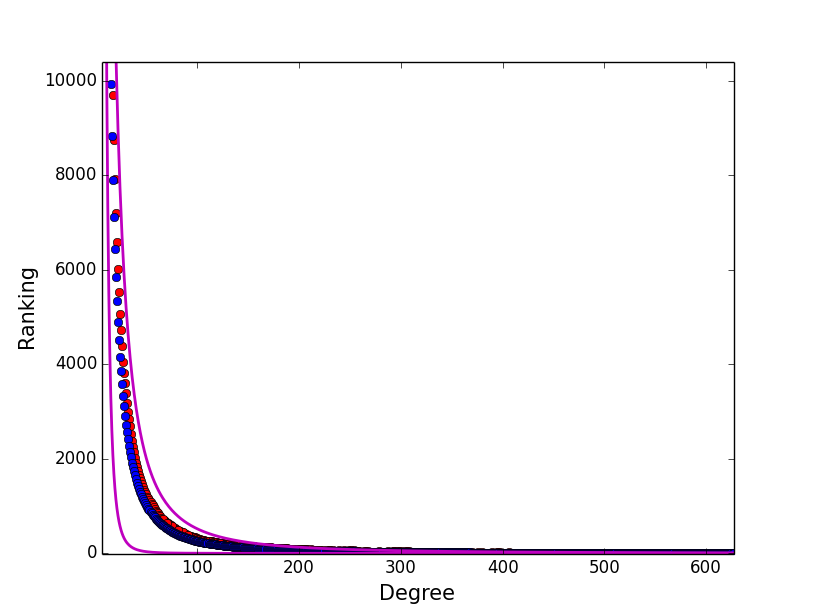}
        }%
        \subfigure[]{%
            \label{fig:fourth}
            \includegraphics[width=0.5\textwidth]{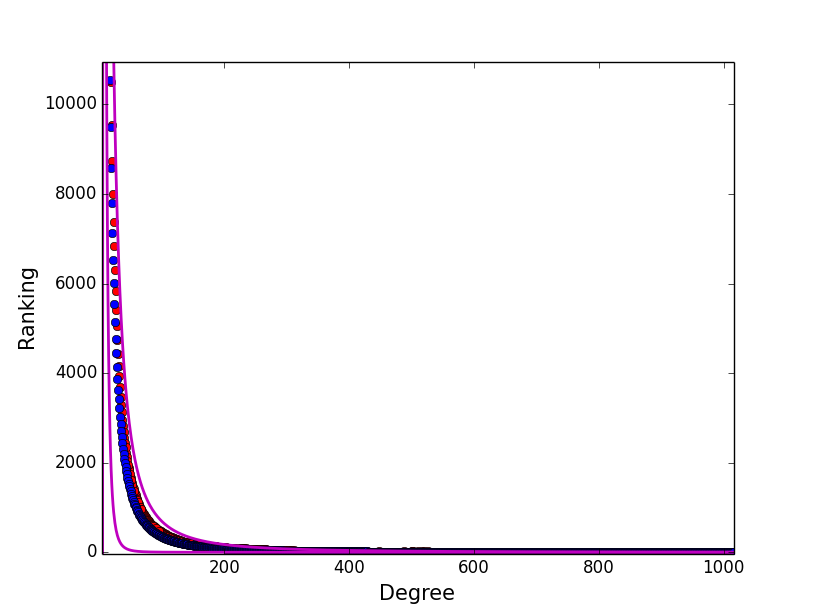}
        }\\ 
          \subfigure[]{%
            \label{fig:fifth}
            \includegraphics[width=0.5\textwidth]{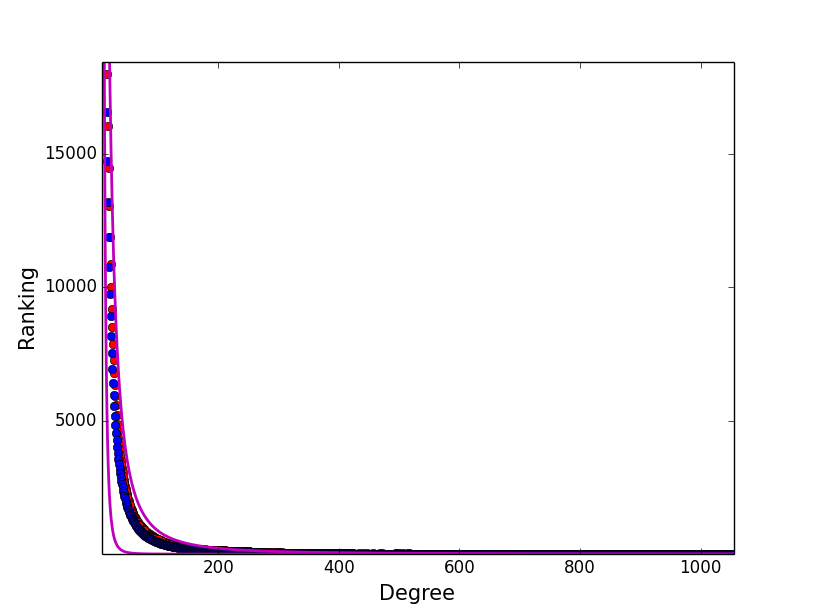}
        }
    \end{center}
    \caption{Plots of actual and expected ranking. Red and blue color dots show actual and expected ranking of a node, and purple thick lines show variance in the ranking as proposed by our method. All results are plotted on BA networks having a)10000 b)30000 c)50000 d)70000 e)90000 nodes, and each new node makes 10 connections when it joins the network. }
   \label{fig:subfigures}
\end{figure}

\begin{table}[ht!]
\centering
\caption{Error in the Actual and Expected Ranking}
\label{my-label1}
\begin{tabular}{|l|l|l|l|l|}
\hline
Number of Nodes & Average Error & Standard Deviation \\ \hline
10000 & 530.80 & 466.67 \\ \hline
30000 & 1613.35 & 1400.92 \\ \hline
50000 & 2684.37 & 2355.16 \\ \hline
70000 & 3782.58 & 3304.09 \\ \hline
90000 & 4860.14 & 4254.90 \\ \hline
\end{tabular}
\end{table}

\section{Conclusion and Future Work}

In this work, a method has been proposed to estimate the degree centrality rank of a node without having a complete structure of the network. The required parameters, such as Network size, minimum, and average degree of the nodes, are computed using the sampling technique. These parameters are used to calculate the power-law exponent of the degree distribution of the network. After getting these parameters, the proposed method can be used to estimate the degree rank of the node. This work can also be used to compare degree ranking of few nodes without calculating the ranking of all nodes. 


\bibliographystyle{unsrt}
\bibliography{mybib.bib}

\begin{thebibliography}{10}

\bibitem{milgram1967small}
Stanley Milgram.
\newblock The small world problem.
\newblock {\em Psychology today}, 2(1):60--67, 1967.

\bibitem{travers1969experimental}
Jeffrey Travers and Stanley Milgram.
\newblock An experimental study of the small world problem.
\newblock {\em Sociometry}, pages 425--443, 1969.

\bibitem{newman2000models}
Mark~EJ Newman.
\newblock Models of the small world.
\newblock {\em Journal of Statistical Physics}, 101(3-4):819--841, 2000.

\bibitem{barabasi1999emergence}
Albert-L{\'a}szl{\'o} Barab{\'a}si and R{\'e}ka Albert.
\newblock Emergence of scaling in random networks.
\newblock {\em science}, 286(5439):509--512, 1999.

\bibitem{caldarelli2007scale}
Guido Caldarelli.
\newblock Scale-free networks: complex webs in nature and technology.
\newblock {\em OUP Catalogue}, 2007.

\bibitem{newman2001clustering}
Mark~EJ Newman.
\newblock Clustering and preferential attachment in growing networks.
\newblock {\em Physical Review E}, 64(2):025102, 2001.

\bibitem{newman2003structure}
Mark~EJ Newman.
\newblock The structure and function of complex networks.
\newblock {\em SIAM review}, 45(2):167--256, 2003.

\bibitem{klemm2002highly}
Konstantin Klemm and Victor~M Eguiluz.
\newblock Highly clustered scale-free networks.
\newblock {\em Physical Review E}, 65(3):036123, 2002.

\bibitem{shaw1954some}
Marvin~E Shaw.
\newblock Some effects of unequal distribution of information upon group
  performance in various communication nets.
\newblock {\em Journal of abnormal and social psychology}, 49(4):547--553,
  1954.

\bibitem{chen2012identifying}
Duanbing Chen, Linyuan L{\"u}, Ming-Sheng Shang, Yi-Cheng Zhang, and Tao Zhou.
\newblock Identifying influential nodes in complex networks.
\newblock {\em Physica a: Statistical mechanics and its applications},
  391(4):1777--1787, 2012.

\bibitem{sabidussi1966centrality}
Gert Sabidussi.
\newblock The centrality index of a graph.
\newblock {\em Psychometrika}, 31(4):581--603, 1966.

\bibitem{freeman1977set}
Linton~C Freeman.
\newblock A set of measures of centrality based on betweenness.
\newblock {\em Sociometry}, pages 35--41, 1977.

\bibitem{stephenson1989rethinking}
Karen Stephenson and Marvin Zelen.
\newblock Rethinking centrality: Methods and examples.
\newblock {\em Social Networks}, 11(1):1--37, 1989.

\bibitem{katz1953new}
Leo Katz.
\newblock A new status index derived from sociometric analysis.
\newblock {\em Psychometrika}, 18(1):39--43, 1953.

\bibitem{sariyuce2013incremental}
Ahmet~Erdem Sariyuce, Kamer Kaya, Erik Saule, and Umit~V Catalyurek.
\newblock Incremental algorithms for network management and analysis based on
  closeness centrality.
\newblock {\em arXiv preprint arXiv:1303.0422}, 2013.

\bibitem{yen2013efficient}
Chia-Chen Yen, Mi-Yen Yeh, and Ming-Syan Chen.
\newblock An efficient approach to updating closeness centrality and average
  path length in dynamic networks.
\newblock In {\em Data Mining (ICDM), 2013 IEEE 13th International Conference
  on}, pages 867--876. IEEE, 2013.

\bibitem{kas2013incremental}
Miray Kas, Matthew Wachs, Kathleen~M Carley, and L~Richard Carley.
\newblock Incremental algorithm for updating betweenness centrality in
  dynamically growing networks.
\newblock In {\em Proceedings of the 2013 IEEE/ACM International Conference on
  Advances in Social Networks Analysis and Mining}, pages 33--40. ACM, 2013.

\bibitem{green2012fast}
Oded Green, Robert McColl, David Bader, et~al.
\newblock A fast algorithm for streaming betweenness centrality.
\newblock In {\em Privacy, Security, Risk and Trust (PASSAT), 2012
  International Conference on and 2012 International Confernece on Social
  Computing (SocialCom)}, pages 11--20. IEEE, 2012.

\bibitem{lee2012qube}
Min-Joong Lee, Jungmin Lee, Jaimie~Yejean Park, Ryan~Hyun Choi, and Chin-Wan
  Chung.
\newblock Qube: a quick algorithm for updating betweenness centrality.
\newblock In {\em Proceedings of the 21st international conference on World
  Wide Web}, pages 351--360. ACM, 2012.

\bibitem{goel2014faster}
Keshav Goel, Rishi~Ranjan Singh, Sudarshan Iyengar, and Sukrit Gupta.
\newblock A faster algorithm to update betweenness centrality after node
  alteration.
\newblock {\em Internet Mathematics}, (just-accepted):00--00, 2014.

\bibitem{nasre2014betweenness}
Meghana Nasre, Matteo Pontecorvi, and Vijaya Ramachandran.
\newblock Betweenness centrality--incremental and faster.
\newblock In {\em Mathematical Foundations of Computer Science 2014}, pages
  577--588. Springer, 2014.

\bibitem{desikan2005incremental}
Prasanna Desikan, Nishith Pathak, Jaideep Srivastava, and Vipin Kumar.
\newblock Incremental page rank computation on evolving graphs.
\newblock In {\em Special interest tracks and posters of the 14th international
  conference on World Wide Web}, pages 1094--1095. ACM, 2005.

\bibitem{bahmani2012pagerank}
Bahman Bahmani, Ravi Kumar, Mohammad Mahdian, and Eli Upfal.
\newblock Pagerank on an evolving graph.
\newblock In {\em Proceedings of the 18th ACM SIGKDD international conference
  on Knowledge discovery and data mining}, pages 24--32. ACM, 2012.

\bibitem{berberich2007comparing}
Klaus Berberich, Srikanta Bedathur, Gerhard Weikum, and Michalis Vazirgiannis.
\newblock Comparing apples and oranges: normalized pagerank for evolving
  graphs.
\newblock In {\em Proceedings of the 16th international conference on World
  Wide Web}, pages 1145--1146. ACM, 2007.

\bibitem{bader2007approximating}
David~A Bader, Shiva Kintali, Kamesh Madduri, and Milena Mihail.
\newblock Approximating betweenness centrality.
\newblock In {\em Algorithms and Models for the Web-Graph}, pages 124--137.
  Springer, 2007.

\bibitem{eppstein2004fast}
David Eppstein and Joseph Wang.
\newblock Fast approximation of centrality.
\newblock {\em J. Graph Algorithms Appl.}, 8:39--45, 2004.

\bibitem{geisberger2008better}
Robert Geisberger, Peter Sanders, and Dominik Schultes.
\newblock Better approximation of betweenness centrality.
\newblock In {\em ALENEX}, pages 90--100. SIAM, 2008.

\bibitem{lee2006statistical}
Sang~Hoon Lee, Pan-Jun Kim, and Hawoong Jeong.
\newblock Statistical properties of sampled networks.
\newblock {\em Physical Review E}, 73(1):016102, 2006.

\bibitem{goldreich2008approximating}
Oded Goldreich and Dana Ron.
\newblock Approximating average parameters of graphs.
\newblock {\em Random Structures \& Algorithms}, 32(4):473--493, 2008.

\bibitem{lovasz1993random}
L{\'a}szl{\'o} Lov{\'a}sz.
\newblock Random walks on graphs: A survey.
\newblock {\em Combinatorics, Paul erdos is eighty}, 2(1):1--46, 1993.

\bibitem{cooper2009random}
Colin Cooper and Alan Frieze.
\newblock Random walks on random graphs.
\newblock In {\em Nano-Net}, pages 95--106. Springer, 2009.

\bibitem{kurant2012graph}
Maciej Kurant, Carter~T Butts, and Athina Markopoulou.
\newblock Graph size estimation.
\newblock {\em arXiv preprint arXiv:1210.0460}, 2012.

\bibitem{hardiman2013estimating}
Stephen~J Hardiman and Liran Katzir.
\newblock Estimating clustering coefficients and size of social networks via
  random walk.
\newblock In {\em Proceedings of the 22nd international conference on World
  Wide Web}, pages 539--550. International World Wide Web Conferences Steering
  Committee, 2013.

\bibitem{katzir2011estimating}
Liran Katzir, Edo Liberty, and Oren Somekh.
\newblock Estimating sizes of social networks via biased sampling.
\newblock In {\em Proceedings of the 20th international conference on World
  wide web}, pages 597--606. ACM, 2011.

\bibitem{cooper2014estimating}
Colin Cooper, Tomasz Radzik, and Yiannis Siantos.
\newblock Estimating network parameters using random walks.
\newblock {\em Social Network Analysis and Mining}, 4(1):1--19, 2014.

\bibitem{dasgupta2014estimating}
Anirban Dasgupta, Ravi Kumar, and Tamas Sarlos.
\newblock On estimating the average degree.
\newblock In {\em Proceedings of the 23rd international conference on World
  wide web}, pages 795--806. ACM, 2014.

\bibitem{fortunato2005make}
Santo Fortunato, Marian Boguna, Alessandro Flammini, and Filippo Menczer.
\newblock How to make the top ten: Approximating pagerank from in-degree.
\newblock {\em arXiv preprint cs/0511016}, 2005.

\bibitem{saxena2017global}
Akrati Saxena and SRS Iyengar.
\newblock Global rank estimation.
\newblock {\em arXiv preprint arXiv:1710.11341}, 2017.

\bibitem{saxena2017observe}
Akrati Saxena, Ralucca Gera, and SRS Iyengar.
\newblock Observe locally rank globally.
\newblock In {\em Proceedings of the 2017 IEEE/ACM International Conference on
  Advances in Social Networks Analysis and Mining 2017}, pages 139--144. ACM,
  2017.

\bibitem{saxena2018estimating}
Akrati Saxena, Ralucca Gera, and SRS Iyengar.
\newblock Estimating degree rank in complex networks.
\newblock {\em Social Network Analysis and Mining}, 8(1):42, 2018.

\bibitem{saxena2017fast}
Akrati Saxena, Ralucca Gera, and SRS Iyengar.
\newblock Fast estimation of closeness centrality ranking.
\newblock In {\em Proceedings of the 2017 IEEE/ACM International Conference on
  Advances in Social Networks Analysis and Mining 2017}, pages 80--85. ACM,
  2017.

\bibitem{saxena2019heuristic}
Akrati Saxena, Ralucca Gera, and SRS Iyengar.
\newblock A heuristic approach to estimate nodes’ closeness rank using the
  properties of real world networks.
\newblock {\em Social Network Analysis and Mining}, 9(1):3, 2019.

\bibitem{saxena2018k}
Akrati Saxena and SRS Iyengar.
\newblock K-shell rank analysis using local information.
\newblock In {\em International Conference on Computational Social Networks},
  pages 198--210. Springer, 2018.

\end{thebibliography}

\end{document}